\documentclass[iop]{emulateapj}

\usepackage{natbib}
\usepackage{amsmath}

\newcommand{\myemail}{billylquarles@gmail.com}

\shorttitle{Stability in $\alpha$ Cen AB}
\shortauthors{Quarles \& Lissauer}

\begin{document}


\title{Long-Term Stability of Planets in the $\alpha$ Centauri System}


\author{B. Quarles\altaffilmark{1}$^,$\altaffilmark{2} and Jack J. Lissauer}
\affil{NASA Ames Research Center, Space Science and Astrobiology Division MS 245-3,
    Moffett Field, CA 94035}
\email{\myemail}

\altaffiltext{1}{NASA Postdoctoral Fellow}
\altaffiltext{2}{Present address: Department of Physics and Physical Science, The University of Nebraska
at Kearney, Kearney, NE 68849}


\begin{abstract}
  We evaluate the extent of the regions within the $\alpha$ Centauri AB star system where small planets are able to orbit for billion-year timescales, and we calculate the positions on the sky plane where planets on stable orbits about either stellar component may appear.  We confirm the qualitative results of \citeauthor{Wiegert1997} (AJ 113, 1445, 1997) regarding the approximate size of the regions of stable orbits, which are larger for retrograde orbits relative to the binary than for prograde orbits. Additionally, we find that mean motion resonances with the binary orbit leave an imprint on the limits of orbital stability, and the effects of the Lidov-Kozai mechanism are also readily apparent.  
\end{abstract}


\keywords{}



\section{Introduction}

The $\alpha$ Centauri triple star system is the nearest neighbor to our Solar System.  The two largest stars in the system, $\alpha$ Cen A and $\alpha$ Cen B, travel about one another on an eccentric orbit with a periapsis of $\sim$11 AU. Both of these stars are {broadly} similar to the Sun in mass and luminosity.  Planetary accretion models suggest that circumstellar planets could have formed within the $\alpha$ Cen system \citep{Quintana2002,Quintana2007}, provided the collision velocities of late stage planetesimals are not too large \citep{Thebault2008,Thebault2009,Thebault2014}.  {\cite{Benest1988} examined the stability of planetary orbits in the $\alpha$ Cen AB system and \citet[henceforth referred to as WH97]{Wiegert1997} performed a more intensive study.}  

Interest in the system has increased in the past few years because of the RV discovery of the close-in planet $\alpha$ Cen B b \citep{Dumusque2012}, the tentative detection of a transiting planet on a somewhat more distant orbit by \cite{Demory2015}, and the possibility of a space mission to attempt to image any planets that may orbit within the habitable zones of $\alpha$ Cen A or $\alpha$ Cen B \citep{Belikov2015a,Belikov2015b,Bendek2015}.  Moreover, a ground-based radial velocity campaign has ruled out the presence of very massive close-in planets \citep{Endl2015} and \citet{Plavchan2015} have shown that a $<$3.3 Earth mass planet in a $\sim$3 day orbit (assuming that the \citet{Dumusque2012} discovery is robust) would be stable when including the effects of tides and General Relativity.  These factors, combined with the advances in computing hardware that enable much larger numerical simulations, motivate us to build upon the work of WH97 and perform long-duration integrations over a high-resolution grid in initial orbital parameters to determine the limits to the regions in this system where planetary orbits are stable on gigayear timescales. 

{Our study of the $\alpha$ Centauri system more precisely identifies the possible regions of parameter space where stable orbits could reside and presents the implications for observing strategies in the search for exoplanets there.  Our methods are outlined in Section 2.  The results of our study are presented in Section 3.  We provide the conclusions of our work and compare our results with previous studies in Section 4.}

\section{Methodology}

The numerical simulations in this paper use a custom version of the \texttt{mercury6} integration package that is designed to efficiently integrate orbits within a binary star system \citep{Chambers2002}, in order to evaluate the long-term stability of planets orbiting stars within the $\alpha$ Centauri system.  The simulations model planets as (massless) test bodies, and integrate each trajectory until a termination event occurs, which can be a collision with either star, a body is ejected from the system, or a specified time interval elapses. Following WH97, we do not consider the effects of tidal or General Relativistic interactions, which are extremely small except very close to each star.  We measure the inclination of the orbits of our test bodies relative to the binary plane.  

Most of our integrations follow the basic model setup of WH97 in terms of a parameter space that includes the semimajor axes of the test particles (initially on near circular orbits) and their inclination relative to the orbital plane of the binary in both circumstellar and circumbinary configurations. For circumstellar planets, we also investigate an alternate parameter space that focuses on test particles that orbit near the plane of the stars (initial inclination $i_o=10^{-6}$ deg.) over a wide range of initial semimajor axes and eccentricities.  

All of our simulations use the nominal values for the stellar masses and the configuration of the binary orbit given in the observational ephemeris derived by \cite{Pourbaix2002}, which we list in Table 1. We also report the uncertainties given in \cite{Pourbaix2002} for completeness.  Our simulations begin the stellar orbit at a mean anomaly of 209.6901$^\circ$, which corresponds to an epoch of JD 2452276.  When considering how the orbits are viewed on the plane of the sky, we apply the necessary transformations to project the binary orbit as well as the orbits of our test bodies into an observer's reference frame.  

We begin the test bodies with their argument of periastron $\omega$, longitude of ascending node $\Omega$, and mean anomaly $M$ taken to be initially zero.  As such, the initial phase of the test particles are not collinear with the binary companion of the star that they are orbiting.  For example, the test bodies orbiting $\alpha$ Cen A (centered at the origin) begin along the positive $y$-axis of the sky plane (North), while $\alpha$ Cen B begins at a point 62.8224$^\circ$ West of North in a Cartesian coordinate system. (See \citet{Deitrick2015} for differences between dynamicist and observer conventions.)

\subsection{Planets on Circumstellar (S-type) Orbits}

We study particles orbiting $\alpha$ Cen A or $\alpha$ Cen B over a range of semimajor axes (0.2 -- 6.0 AU), as motivated from the previous study by WH97 and from the statistical fitting formula developed for planets within binaries \citep{Holman1999}.  In contrast to WH97, we use the hybrid symplectic method for integration with \texttt{mercury6}.  This allows us to be more uniform with the choice of the starting timestep while maintaining a tight control on the errors in energy and angular momentum.  We employ a timestep of 0.005 yr step$^{-1}$ for all our circumstellar runs to properly handle moderately close approaches with the host star.

We analyze two separate regions of parameter space to estimate the stability of test bodies that interact through gravitational forces from the stars.  In each region, we evaluate a grid of initial conditions across a range of semimajor axes considering either initially circular orbits with increments in the mutual inclination relative to the binary plane of motion or planar orbits with a range of eccentricities for the test body.   

The mutual inclination regime ($a,i$) follows the methodology of WH97, where the test bodies begin on nearly circular ($e_o=10^{-6}$) circumstellar orbits starting with a semimajor axis from 0.2 AU to 6.0 AU with an increment of 0.01 AU relative to the host star.  We evaluate 581 test particles per degree of mutual inclination, resulting in a grid of (181$\times$581) different initial conditions that provide a high-resolution view of the system.  Within this parameter space, two characterizations of motion develop that are described as either prograde or retrograde.  In our usage these terms refer to the direction of orbital motion relative to the binary orbit, i.e., a prograde particle orbits its star in the same direction as the hosting star does around the center of mass.  Prograde and retrograde classifications are assigned by the starting inclination of the test particle, where starting values of $i$ less than 90$^\circ$ correspond to prograde and retrograde orbits begin with values of $i$ greater than 90$^\circ$.  Although retrograde objects exist within the Solar System, they are expected to have undergone a more complicated early evolution with gas disk interactions \citep{Jewitt2007}, higher-order three-body effects \citep{Lithwick2011,Naoz2011}, and even possibly have arisen from a more drastic shift in the architecture of the Solar System giant planets \citep{Gomes2005,Tsiganis2005}.  Thus, we may consider the retrograde cases to be less likely, but include them in our analyses for a more complete picture.

The second parameter space ($a,e$) prescribes the starting values of eccentricity and semimajor axis of the test bodies and assumes that they begin with a small (10$^{-6}$ degree) inclination relative to the binary to allow for the freedom of motion in the third dimension.  The semimajor axis values follow the same range and resolution as before, but the starting eccentricity of each test body ranges from 0.0 to 1.0 with increments of 0.01, producing a grid of (101$\times$581) initial conditions.  The initial configuration of the binary and phase of the test objects in these simulations also follow the prescription detailed in the previous regime.

In order to explore efficiently the transition to instability for our circumstellar runs around $\alpha$ Cen B on timescales up to 1 Gyr, we initially perform our runs for 10 Myr and based on those results we only continue a subset of surviving particles near the stability boundary for the full 1 Gyr.  Approximate symmetry exists between the masses of the stars, so we only consider orbits around $\alpha$ Cen A for 100 Myr.  

\subsection{Planets on Circumbinary (P-type) Orbits}

A different class of planet may be present {that orbits} both stars (circumbinary) in the system of $\alpha$ Cen AB around the barycenter.   We also consider circumbinary test particles within the mutual inclination regime, where we evaluate test bodies from 35 -- 100 AU with an increment of 0.1 AU.  This results in a grid of (181$\times$651) different initial conditions.  For these integrations, we use a timestep of 2 yr step$^{-1}$ which was adequate to control the numerical error. The computational cost of circumbinary orbits is quite low, which allows us to evaluate all of the (surviving) test particles in that scenario for the full 1 Gyr.

\section{Results}

\subsection{Inclined, Circular Circumstellar Initial Orbits} \label{sec:inclined}
\begin{figure*}
{
\plotone{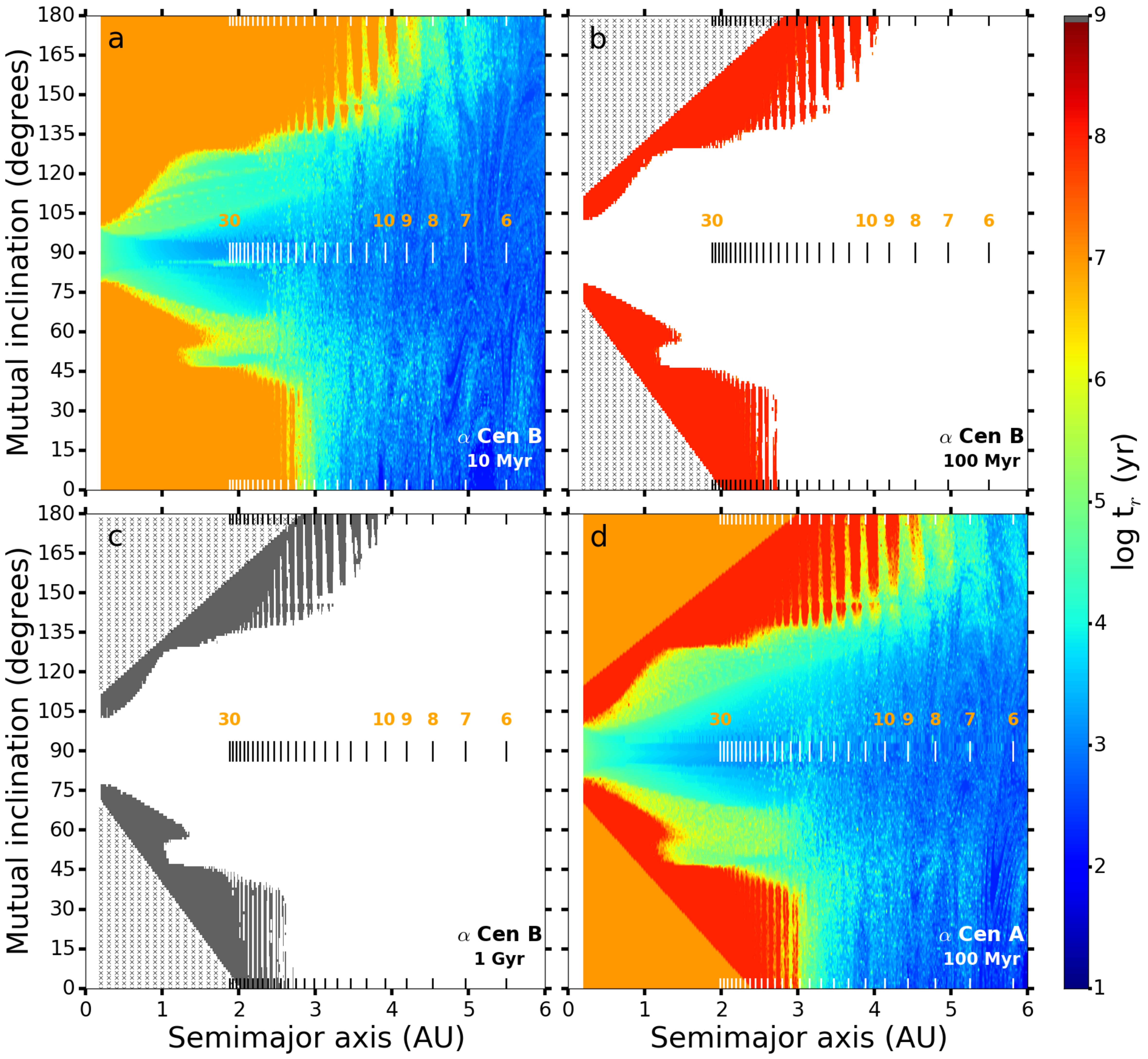}}
\caption{Removal times, $t_r$, (due to collisions/ejections) for test particles starting on nearly circular circumstellar orbits are displayed with respect to the initial semimajor axis and mutual inclination relative to the stars' orbital plane.  The color scale (right) for each panel denotes the removal time for a given test particle and is logarithmic in scale ranging from 10 yr to 10$^9$ yr.  Regions colored orange, red, and dark gray indicate those that are stable for 10 Myr, 100 Myr, and 1 Gyr, respectively.  Panel (a) shows results for all test particles orbiting $\alpha$ Cen B for 10 Myr of simulation time and we provide only the subset of test particles that were further evaluated and survived for the entire simulation timescale of 100 Myr (b) and 1 Gyr (c).  The region in (b) and (c) denoted by ``x'' marks and the orange regions in (d) were excluded from longer simulations, but are expected to be long-term stable.  Panel (d) contains the results when considering test particles orbiting $\alpha$ Cen A up to 100 Myr.  Additionally, the white/black ticks denote the locations of the (internal) $N$:1 mean motion resonances ($6\leq N \leq 30$) for the binary stars. 
 \label{aCen_ia_time}}
\end{figure*}

Figure \ref{aCen_ia_time} shows the removal timescale as a function of the starting parameters of a test object.  Large regions of parameter space become unstable on a relatively short ($\ll$1 Myr) timescale.  The boundary between stable and unstable starting conditions is complex.  Small regions of local instability are present due to $N$:1 mean motion resonant interactions, where $N$ is an integer, with the stellar perturber.  As noted by WH97, the Lidov-Kozai (L-K) mechanism \citep{Lidov1962,Kozai1962} is an important process in destabilizing orbits substantially inclined to the plane of the stellar orbit.  Moreover, our results confirm the general findings of WH97 that the retrograde starting conditions yield a broader region of parameter space with stable orbits than do prograde initial conditions {\citep{Henon1970}}.

Mean motion resonances extend over a range of mutual inclinations and mainly affect particles with semimajor axes larger than 2 AU, whereas the L-K mechanism drives the test particles to high values of eccentricity for the mutual inclinations between the critical values, 39.2$^\circ$ -- 140.8$^\circ$ \citep{Innanen1997}.  Our results reproduce a broad region of instability between 75$^\circ$ -- 105$^\circ$ that was identified previously by WH97 where all test particles are removed on relatively short timescales.  For initial parameters in the prograde ($i<90^\circ$) regime of each panel, the extent of stable test particles decreases with increasing staring inclination.  For the retrograde ($i>90^\circ$) regime, similar features (MMRs and the L-K mechanism) are evident but present themselves differently, with larger starting semimajor axes maintaining stability despite an increased proximity to the stellar companion.  The stability of test particles that orbit $\alpha$ Cen A follow similar trends as those initially orbiting $\alpha$ Cen B, albeit at slightly larger starting semimajor axis.  

\begin{figure*}
\plotone{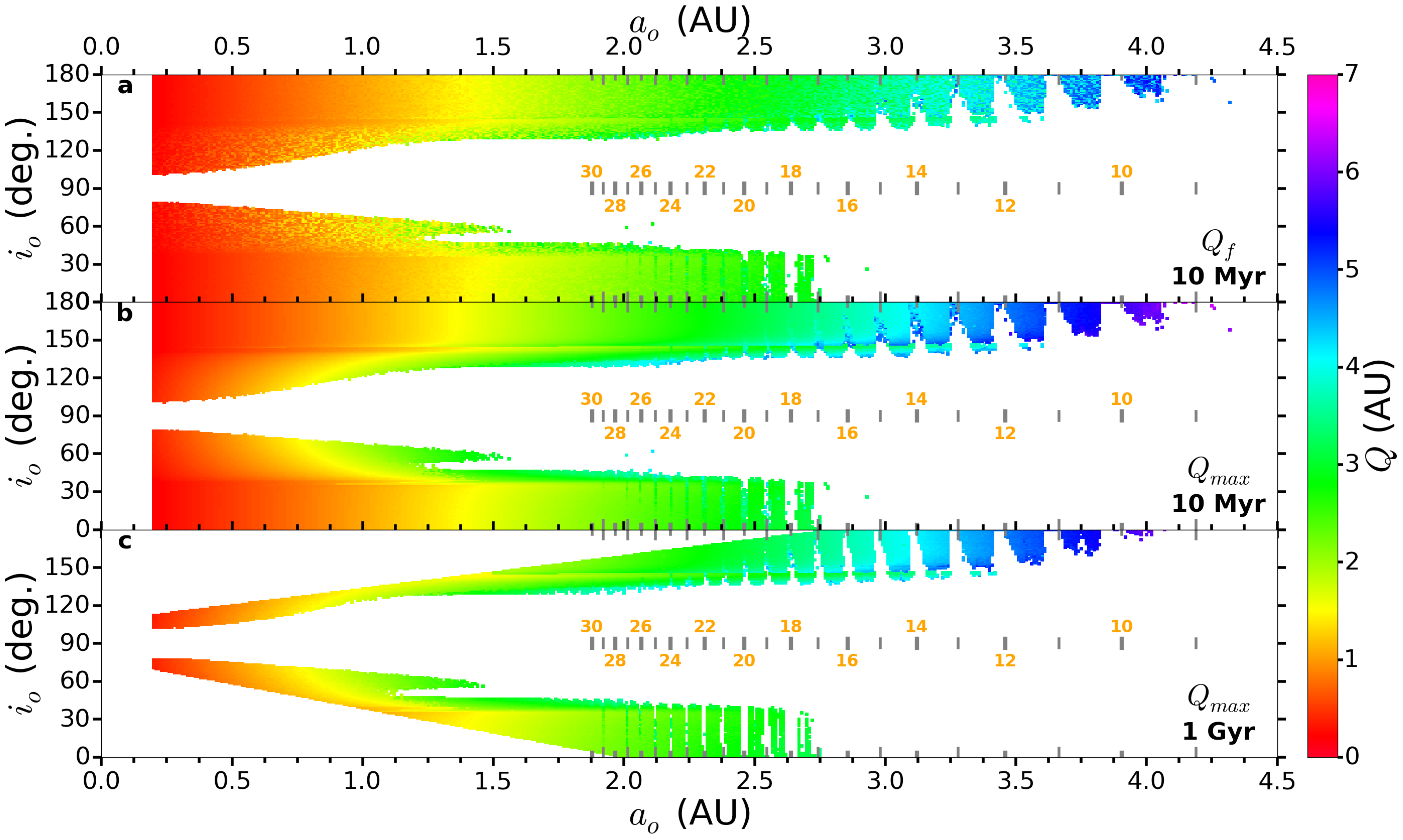}
\caption{Apastron distances, $Q$, of the surviving test particles orbiting $\alpha$ Cen B (Figures \ref{aCen_ia_time}a \& \ref{aCen_ia_time}c) are represented using the color scale shown on the right.  (a) Value of $Q$ at 10 Myr, $Q_f$.  (b) Maximum value of $Q$ obtained during the 10 Myr integrations, $Q_{max}$.  (c) Maximum value of $Q$ obtained by these particles that survived the 1 Gyr integrations, $Q_{max}$.  Note that the trends in $Q_{max}$ and $Q_f$ related to the $N$:1 mean motion resonances (MMRs) and the Lidov-Kozai Mechanism.  The gray ticks denote the $N$:1 MMRs ranging from 9:1 to 30:1. \label{Q_ai}}
\end{figure*}

The results presented in Figure \ref{aCen_ia_time} correlate the initial starting parameters to a characteristic removal time from the system (either by ejection or collision), but the final states are dynamically evolved with larger eccentricities.  We illustrate these evolved states using color-coded maps of the test particles that survive 10 Myr around $\alpha$ Cen A with colors representing the apastron distance, $Q$ (Figure \ref{Q_ai}).  The apastron distance evolves with time, so we show maps of both the final apastron values, $Q_f$, attained over the course of a simulation and the maximum values, $Q_{max}$.  The values of $Q_f$ illustrate the ``phase mixing'' of the oscillating eccentricities and give an estimate of likely distances that can be achieved at a given epoch.  

Most of the same spatial structures are present in Fig. \ref{Q_ai}a as in Fig. \ref{aCen_ia_time}a, but the color scale demonstrates the differences from the expected smooth gradient of values and the largest distances from the host star that are permitted.  These differences accentuate the two previously mentioned processes that modify the eccentricity of the test particle over a simulation.  For example, there are discontinuities at the border of the L-K regime for both prograde and retrograde orbits.  Between the critical values of mutual inclination, 39.2$^\circ$ -- 140.8$^\circ$, the gradient of $Q_{max}$ is non-linear (Fig. \ref{Q_ai}b), indicating where in the parameter space that the L-K mechanism is effective at increasing the eccentricity.  In the retrograde regime, long-lived test particles can attain large $Q_{max}$ values (up to $\sim$7 AU).  In contrast, Fig. \ref{Q_ai}b illustrates further that values of $Q$ change in time and that nearby initial conditions can exhibit significantly different variations, as shown by the sharp gradients of colors where dynamical processes are active.  Figure \ref{Q_ai}c demonstrates test particles that attain large values of $Q$ ($\gtrsim$5 AU) in the retrograde regime and can remain stable over billion-year timescales.

\begin{figure*}
\plotone{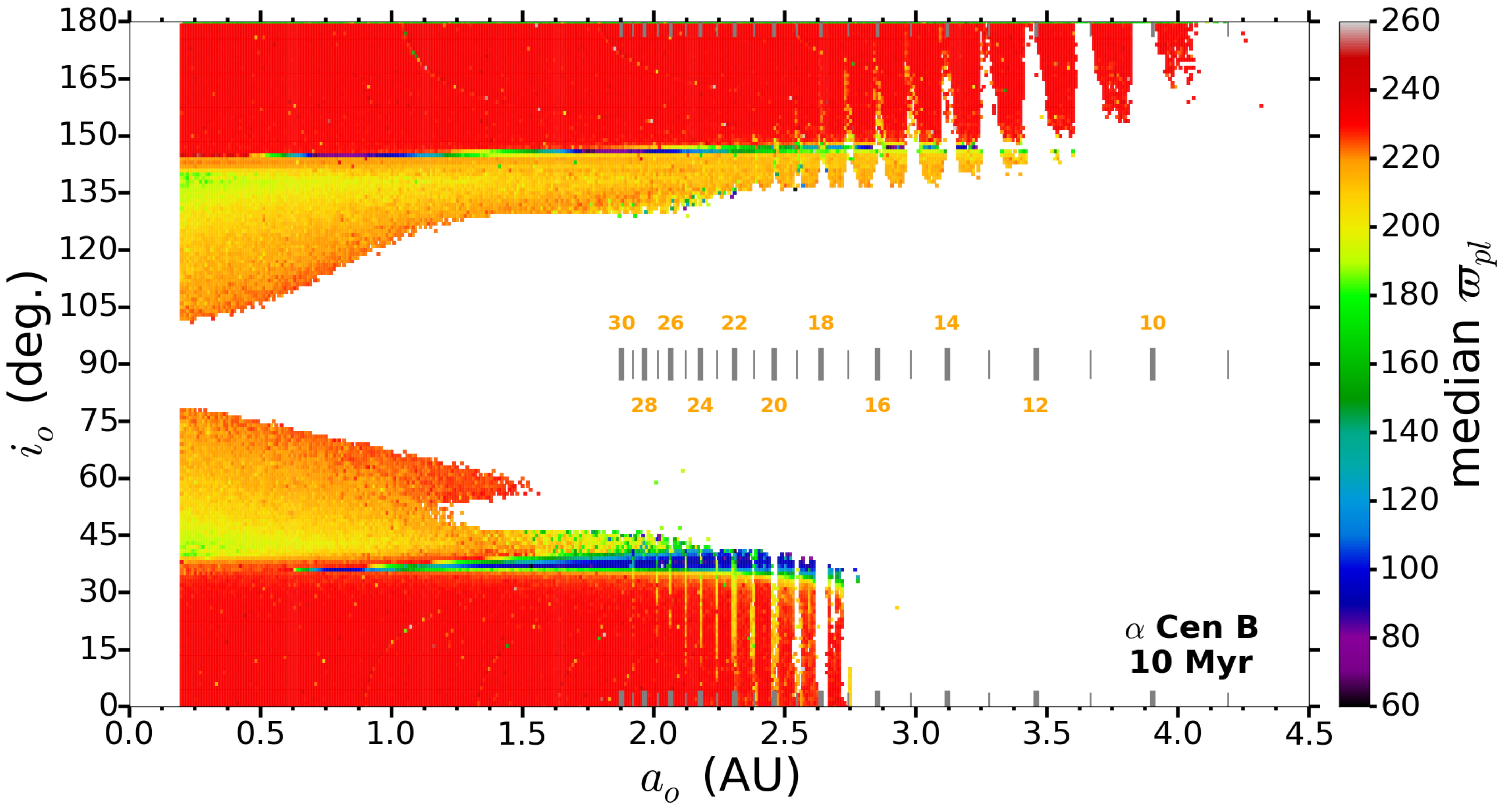}
\caption{The median value of the longitude of periastron ($\varpi_{pl}$) for orbits of the surviving test particles in Figure \ref{aCen_ia_time}a.  This value has been determined using the standard longitude of periastron $\varpi = \omega + \Omega$ for prograde ($i_o<90^\circ$) and $\varpi^\dagger = \omega - \Omega$ for retrograde ($i_o>90^\circ$) orbits.  Note that the binary longitude of periastron $\varpi_\star=231.65^\circ$ and a large number of points are aligned with this value.\label{ai_res}}
\end{figure*}

We probe deeper into the full extent of variations due to MMRs and the L-K mechanism by plotting in Figure \ref{ai_res} the median value of the longitude of periastron, $\varpi = \omega \pm \Omega$, of the surviving test particles after 10 Myr.  The $N$:1 MMRs and the L-K mechanism both involve terms related to the periastron of the orbits that contribute to the calculation of an associated resonant argument, $\phi = \varpi - \varpi_\star$.  We inspect this value (relative to neighboring values) to identify the appearance of possible resonances.  We do not determine specifically whether these resonances are long-lived, only that libration of the resonant argument can occur over a fraction of the time during the evolution of the test particles.  Figure \ref{ai_res} and Figure \ref{aCen_ia_time}c indicate a spatial and dynamical change in stability near the critical inclination for the L-K mechanism.  In regards to the long-term stability, we look at these limiting boundaries of possible resonance and we find that the MMRs can be effective on gigayear timescales up to 30:1 ($\sim$2 AU) for the prograde cases and up to 20:1 ($\sim$2.5 AU) for the retrograde cases.  The boundary for the L-K regime becomes distinct in this view and other variations appear in regions outside the regime for the L-K mechanism, possibly indicative of other secular interactions.  Moreover, a large number of points outside of the L-K regime (both prograde and retrograde) tend to have a longitude of periastron nearly equal to that of the binary orbit.

\subsection{Planar, Eccentric Circumstellar Initial Orbits}
\begin{figure*}
{
\plotone{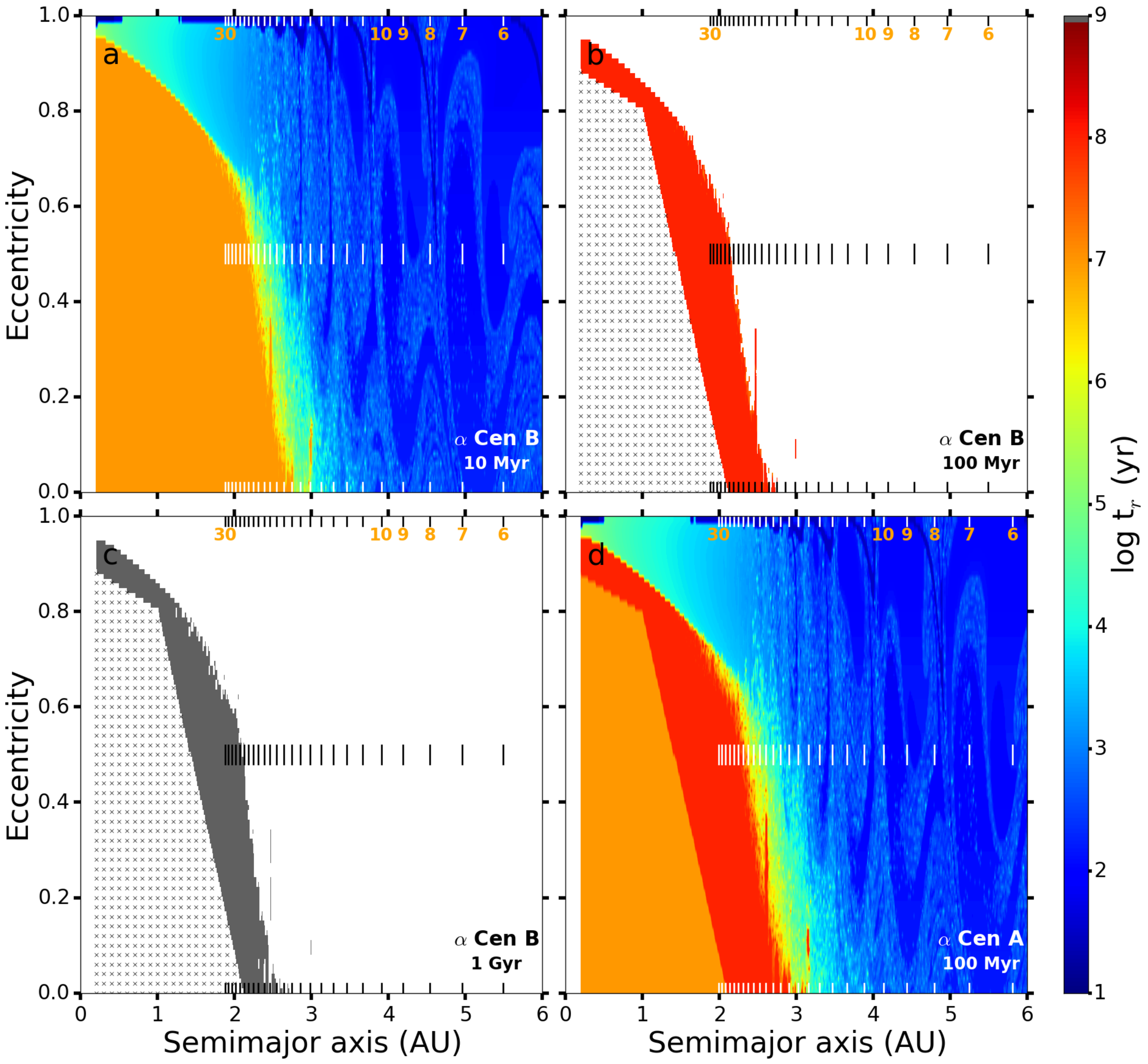}}
\caption{Removal times using the full range of initial eccentricity for the test particles and starting with nearly planar prograde orbits about $\alpha$ Cen B (a, b, \& c) and $\alpha$ Cen A (d).  Each panel shows results for the same timescale as the corresponding panel in Figure \ref{aCen_ia_time}. The color scale (right) for each panel denotes the removal time for a given test particle and is logarithmic in time ranging from 10 yr to 10$^9$ yr.  The region in (b) and (c) denoted by ``x'' marks and the orange regions in (d) were excluded from longer simulations, but are expected to be long-term stable.
\label{aCen_ea_time}}
\end{figure*}

We simulated test bodies across a different portion of parameter space, where the initial states began nearly planar with the binary ($i_o=10^{-6}$ deg.) but a full range of initial eccentricities were allowed.  Figure \ref{aCen_ea_time} shows stability diagrams of this region of parameter space, arranged by timescale and the host star in a similar manner as those in Fig. \ref{aCen_ia_time}.  This region of parameter space shows a sharp boundary between the stable and unstable runs.  Not surprisingly, the maximum stable semimajor axis decreases when the initial eccentricity, $e_o$, of the test body is increased; we find that the rate of decrease is slow for $e_o<0.8$ and rapid for higher eccentricity.  Above a starting eccentricity of 0.8, an actual planet would be affected by other interactions (i.e., tides, oblateness, General Relativity) at periastron $q$ ($= a(1-e)$) and/or significant perturbations from the stellar companion at apastron $Q$.  Although we have not considered the effects of tides or General Relativity and our choice of a constant timestep compromises the accuracy of our integrations when $q$ is small, we see the effects of large perturbations at apastron destabilizing the orbits of most test particles with eccentricities greater than 0.8.

Within this region of parameter space, we probed how the stability boundary varied with respect to the initial eccentricity of the test body by considering the largest stable starting semimajor axis for a given initial eccentricity.  We define these points as the ``stability boundary'' up to the full simulation time.  This procedure was used in previous studies of planetary stability in binary systems \citep{Dvorak1986,Rabl1988,Holman1999}, but they have focused on the eccentricity of the stars.  Using this stability boundary, we found that the variation of the largest stable periastron distance $q$ was approximately linear with respect to the planetary eccentricity (see also \citet{Popova2012}), which allowed us to apply Monte Carlo methods to determine the best fitting parameters for the slope $m$ and $y$-intercept $b$ of a linear function (i.e., $y=mx+b$) within a parameter space of eccentricity and the value of $q$.  We fitted a line to the stable starting values of $q$, using \textit{emcee} \citep{Foreman-Mackey2013}, allowing us to transform into our input variables of semimajor axis and eccentricity via the following:

\begin{equation}
e = mq + b.
\end{equation}

\noindent For low-mass bodies orbiting $\alpha$ Cen A, we find $m=-0.356 \pm 0.012$ and $b=0.913 \pm 0.011$ and are able to produce a fit to the instability boundary through some algebra to obtain

\begin{equation} e = {(-0.356 \pm 0.012)a + 0.913 \pm 0.011\over 1 + (0.356 \mp 0.012)a}. \end{equation}

\noindent The same type of fitting can be applied for test bodies orbiting $\alpha$ Cen B, where we find $m=-0.367 \pm 0.012$ and $b=0.890 \pm 0.012$.  In this case the largest stable semimajor axis for an initially circular orbit ($a_o = -b/m$) is slightly smaller than for bodies that orbit $\alpha$ Cen A.  Also, we note that we find the largest stable semimajor axis for an initially planar, circular prograde orbit around $\alpha$ Cen A to be $\sim$2.56 AU after 100 Myr of simulation time, which is larger than the value of 2.34 AU from WH97 and suggests that the erosion of initial states for this case is not as severe over long integration timescales as previously indicated. 

\begin{figure*}
\plotone{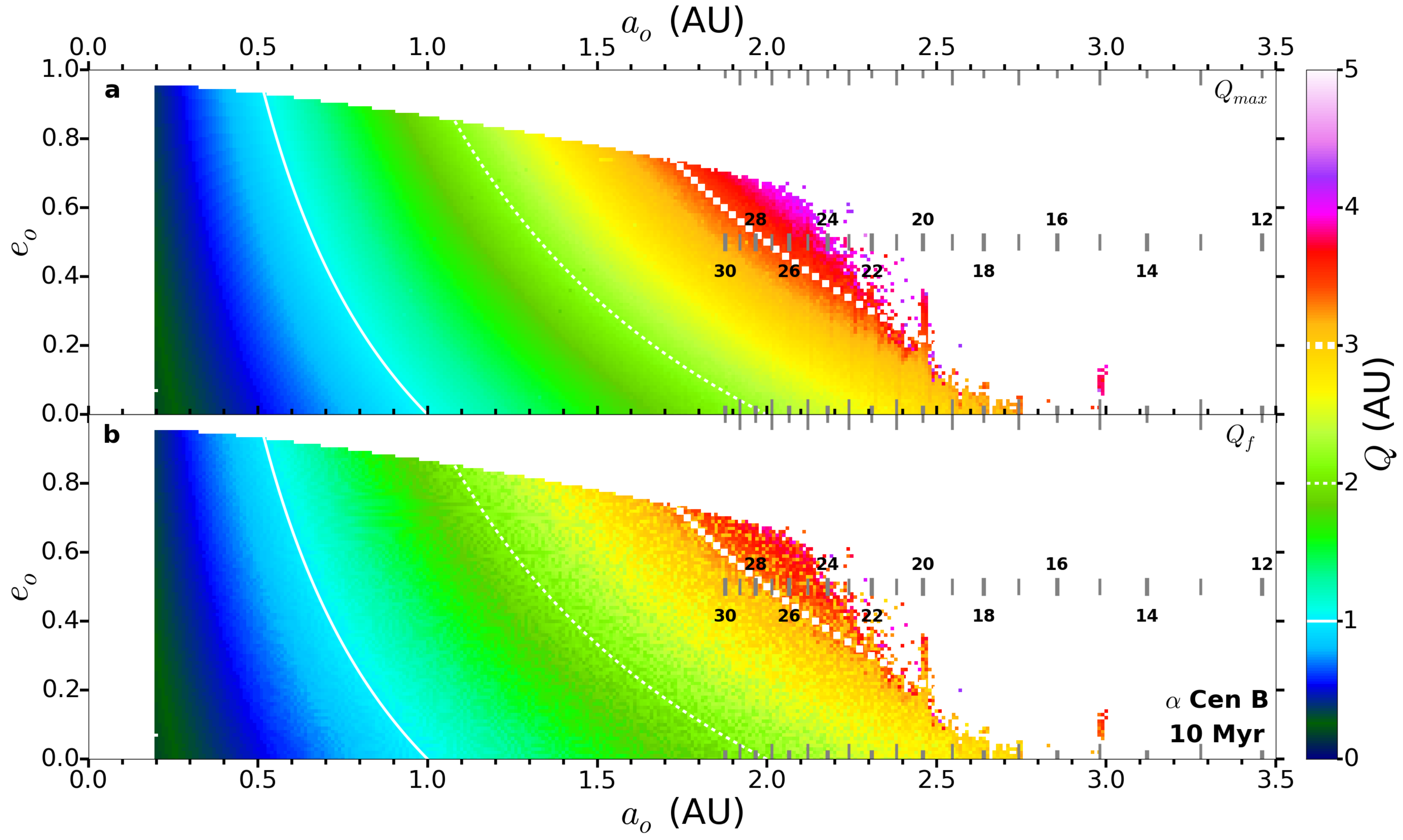}
\caption{Similar to Figure \ref{Q_ai} but using results for long-lived particles shown in Figure \ref{aCen_ea_time}a. There are trends in the maximum $Q_{max}$ (a) and final $Q_f$ (b) related to the $N$:1 mean motion resonances (MMRs) that occur most strongly when $N$=15 or 20.  The color scale indicates the value of $Q$ and elucidates the regions of parameter space that are likely dynamically active.  The gray ticks denote the $N$:1 MMRs ranging from 12:1 to 30:1.  White curves illustrate contours of constant $Q$ at 1 (solid), 2 (dashed), \& 3 (squares) AU. \label{Q_ae}}
\end{figure*}

The removal times do not provide a full dynamical picture of the parameter space, thus we produce views in Figure \ref{Q_ae} illustrating the variations in both the maximum and final values of $Q$.  Figure \ref{Q_ae}a demonstrates that $Q_{max}$ extends smoothly from low to high starting semimajor axes apart from the effects of MMRs near the stability limit.  The 20:1 and 15:1 MMRs produce clusters of ``lucky'' particles that  likely survive for long timescales due to their initial conditions and orbital phase relative to the epoch of binary periastron.  Figure \ref{Q_ae}b shows time-scale variations indicative of oscillation in $Q$ that have become out of phase.  The white curves show that $Q$ changes little for $Q_o<2.5$ AU.  Note that some initial conditions can achieve a value of $Q$ up to $\approx$5 AU.

\subsection{Temporal and Radial Extent}
\begin{figure*}
\epsscale{1.2}
\plotone{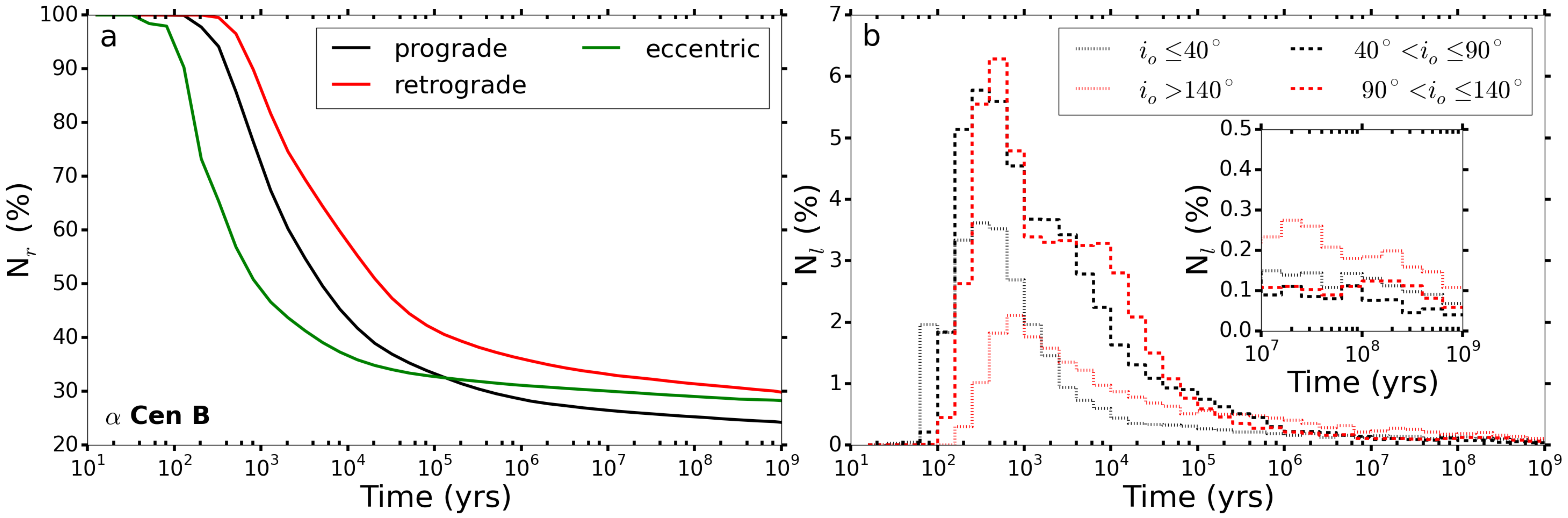}
\caption{Decay curve showing the fraction of surviving test particles relative to the initial populations from the initial eccentric (green), prograde inclined (blue), and retrograde inclined (red) runs that orbit $\alpha$ Cen B.  (a) The percentage of remaining particles (N$_r$) are given in simulation time (yrs) and normalized relative to the appropriate initial population.  (b) The percentage of the initial population that are lost (N$_l$) in the initially prograde and retrograde (near circular) runs are further decomposed into subregions of inclination and logarithmic bins in time of width $10^{1/5}$.  These histograms further illustrate the small fraction ($\sim$1\%) of test particles lost between 10$^7$ and 10$^9$ yrs.  The inset panel in (b) shows a zoomed view of this temporal domain for clarity. \label{N_hist}}
\end{figure*}

We quantify the full extent of our simulations by two measures: the rate of survival/loss as a function of simulation time and the largest distance a bound test particle can achieve over the course of a simulation.  Figure \ref{N_hist} demonstrates how the number of test particles remaining $N_r$ (Fig. \ref{N_hist}a) and the number lost $N_l$ (Fig. \ref{N_hist}b) change with time (logarithmic scale).  These results in Fig. \ref{N_hist} are derived from simulations of test particles initially orbiting $\alpha$ Cen B, and similar results are found when considering $\alpha$ Cen A as the host star.

In Figure \ref{N_hist}a, the results have been grouped into prograde (black), retrograde (red), and eccentric (green) as previously defined in Figs. \ref{aCen_ia_time} and \ref{aCen_ea_time}.  The prograde and retrograde runs each start with 52,290 test particles, whereas the eccentric parameter space begins with 58,681.  Of these initial states, the eccentric simulations retain more particles than the prograde, which indicates that inclination has a more adverse effect on stability than does eccentricity.

Figure \ref{N_hist}b illustrates the time at which test bodies are removed for the initial nearly circular orbit runs, which are divided into prograde and retrograde groupings.  The corresponding histogram for the eccentric runs looks similar, but the main peak is shifted slightly to shorter times.  For each of the ranges of inclination that we examine, the peak in the bodies lost per logarithmic bin in time occurs near 1000 years.  The percentage (of the initial population) lost between 10 and 100 Myr is $\sim$3\%, but only about half as many particles are lost in the next decade (100 Myr -- 1 Gyr).  This amounts to about $\sim$1500 test bodies being lost between 100 Myr and 1 Gyr of simulation time along the transition region of stability in the inclination parameter space, or around 7 test bodies per degree of mutual inclination.  But, differences occur when we consider the magnitude of the peak percentage lost with respect to the L-K regime (dashed) and those outside (dotted) the region.  The percentage lost beyond 100 Myr differs in the retrograde regime for initial inclinations above 140$^\circ$ when compared to those below 140$^\circ$, implying that retrograde test particles near the MMRs become unstable on a longer timescale. 

\begin{figure*}
\plottwo{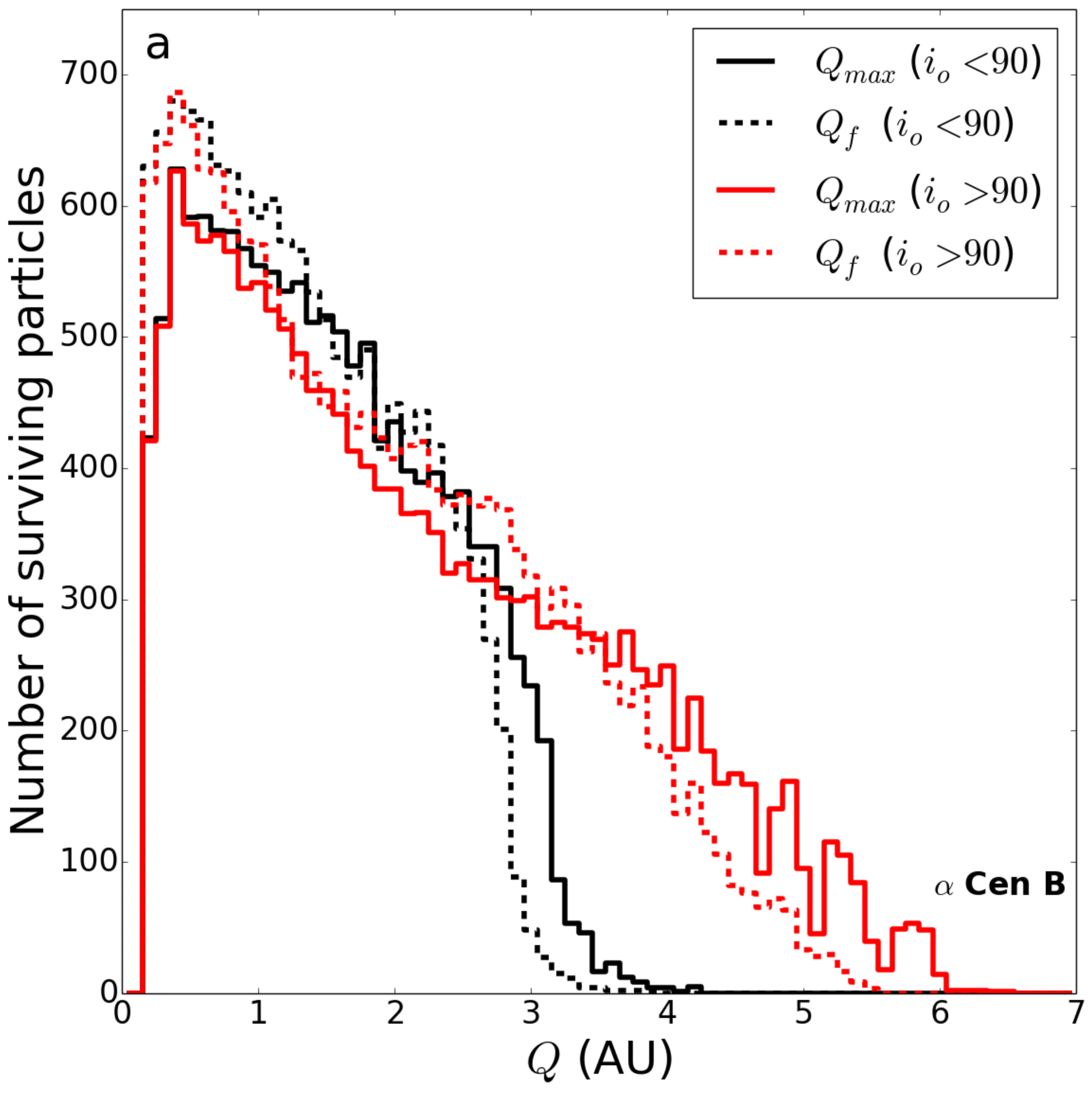}{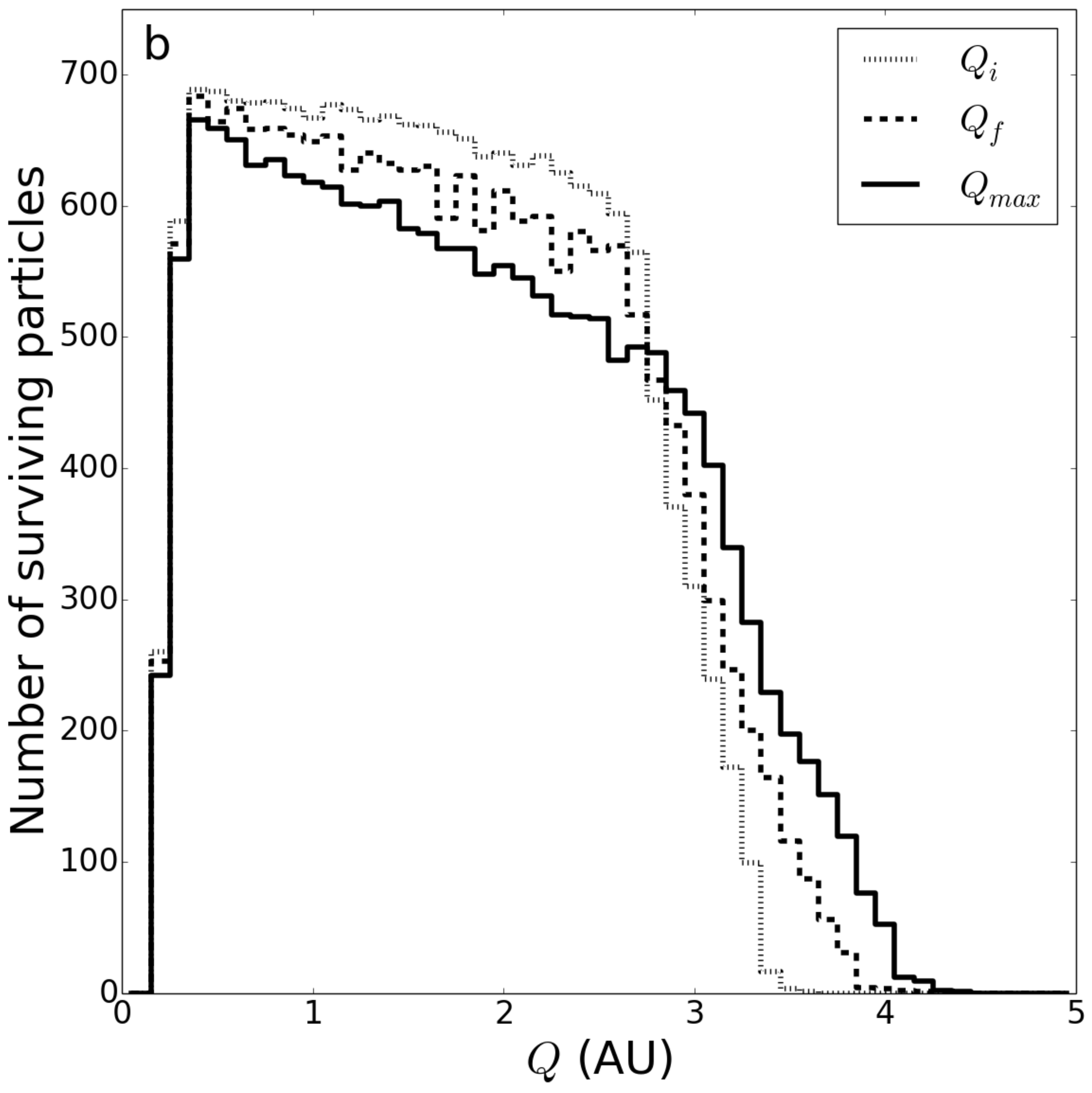}
\caption{The distributions of apastron distance, $Q$, of surviving particles from the 10 Myr simulations around $\alpha$ Cen B.  For the circular inclined runs (a), the distributions extend to substantially higher values of $Q$ for retrograde (red) than prograde (black) particles.  Smaller distinctions are present when comparing $Q_{max}$ (solid) to $Q_f$ (dashed).  The eccentric runs (b) are similar to the prograde circular inclined runs but extend farther, with more particles having $Q\gtrsim 4$ AU.  The initial distribution of the surviving particles ($Q_i$) is included to illustrate the extent of diffusion in $Q$ over 10 Myr of evolution. \label{Q_hist}}
\end{figure*}

Observers would like to know how far from the host star a typical test body can remain stable. Figure \ref{Q_hist} illustrates how the apastron distance $Q$ differs between the prograde (black) and retrograde (red) cases in Fig. \ref{Q_hist}a.  In this view, we see that most of the surviving test particles exist within 3 AU (prograde) and 5 AU (retrograde).  Almost all of the stable prograde test particles remain within 4 AU.  In contrast, the falloff in particle numbers with distance in the retrograde case is more gradual with some stable particles spending time beyond 6 AU and a few reaching a distance of $\sim$7 AU.  For the eccentric runs (Fig. \ref{Q_hist}b), the bulk of surviving particles are within 3 AU as is the case for the inclined, prograde runs; however, the eccentric runs extend $\sim$0.25 AU farther in $Q_f$ and $\sim$0.5 AU in $Q_{max}$.  This becomes important because the extent of the values of $Q$ gives information as to how far away from a host star one should potentially look for circumstellar planets in $\alpha$ Cen.  Also, we note that some of the largest values of $Q_{max}$ represent particles between 2 -- 3 AU, where MMRs aid in the excitation of eccentricity.

\subsection{Circumbinary planets}
Figure \ref{CBP_time} shows our results for planets on circumbinary orbits.  As with circumstellar orbits, the extent of the stable region for retrograde orbits is larger than that for prograde orbits.  However, inclinations around 90$^\circ$ are typically more stable than lower inclinations, in sharp contrast to the trend for  circumstellar case (Section \ref{sec:inclined}).  This is likely due to a slightly different interaction with the quadrupole moment \citep{Cuk2005}, which couples the variation of inclination with the eccentricity of the inner binary.  As a result, initial conditions between $\sim$$40^\circ - 140^\circ$ of mutual inclination can periodically switch from prograde to retrograde.  Also, the effects of external MMRs, which can cause instabilities through chaos and resonance overlap \citep{Mudryk2006,Wisdom1980,Chirikov1979}, are clearly evident in Figure \ref{CBP_time}.

\subsection{Maps for Observers}
\begin{figure*}
{
\plotone{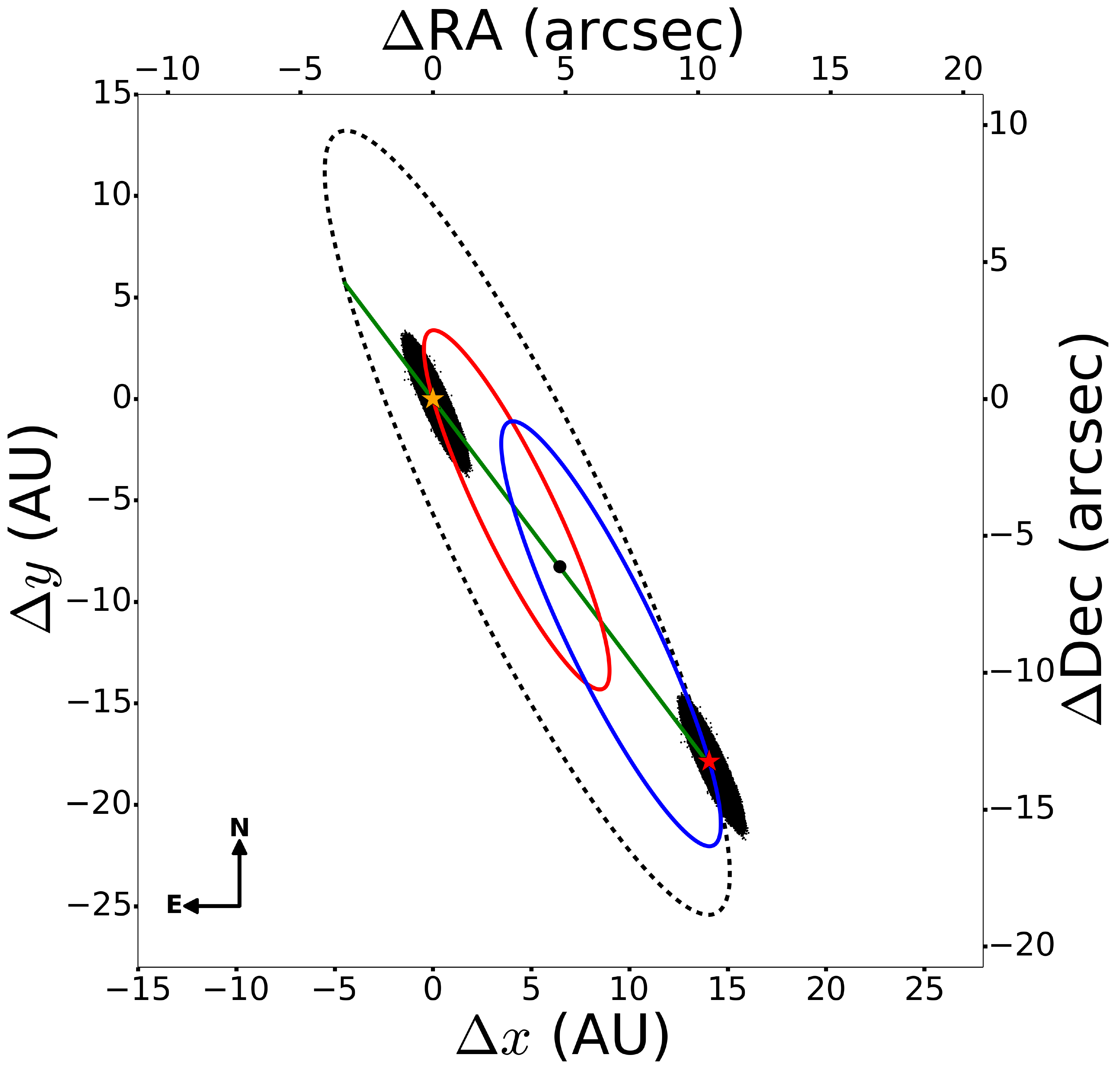}}
\caption{Projection of the stellar orbit of $\alpha$ Cen AB at apastron onto the sky.  The astrocentric orbit $\alpha$ Cen B about $\alpha$ Cen A is shown by the dashed curve, with the stars shown at apastron.  The center of mass (dot near the center of the image) is shown along with the barycentric ellipses for both $\alpha$ Cen A (red) and $\alpha$ Cen B (blue).  Disks have been placed around each star to illustrate the areal coverage of stable test particles orbiting with the plane of the binary.  The scale for potential observers in Right Ascension and Declination is given on the top and right axis, respectively.  \label{sky_bin}}
\end{figure*}

We have illustrated the regions of phase space where planets may reside within the $\alpha$ Centauri system, but to find such worlds we also consider how their orbits appear to an observer in our Solar System through a projection onto the sky plane.  Planets on circumbinary orbits could appear anywhere in the sky in the general vicinity of  $\alpha$ Cen, but those on  circumstellar orbits would only appear close to their stellar host.  Figure \ref{sky_bin} shows the sky projected view of the binary orbit in both the astrocentric (dashed) and barycentric (red \& blue) coordinate frames.  Because this is a projected view, we note that the points of binary apastron and periastron do not fall on the extreme points of the dashed line.  Thus, we've included a projection of the major axis for the astrocentric frame centered on $\alpha$ Cen A. 

\begin{figure*}
\centering
\includegraphics[scale=0.18]{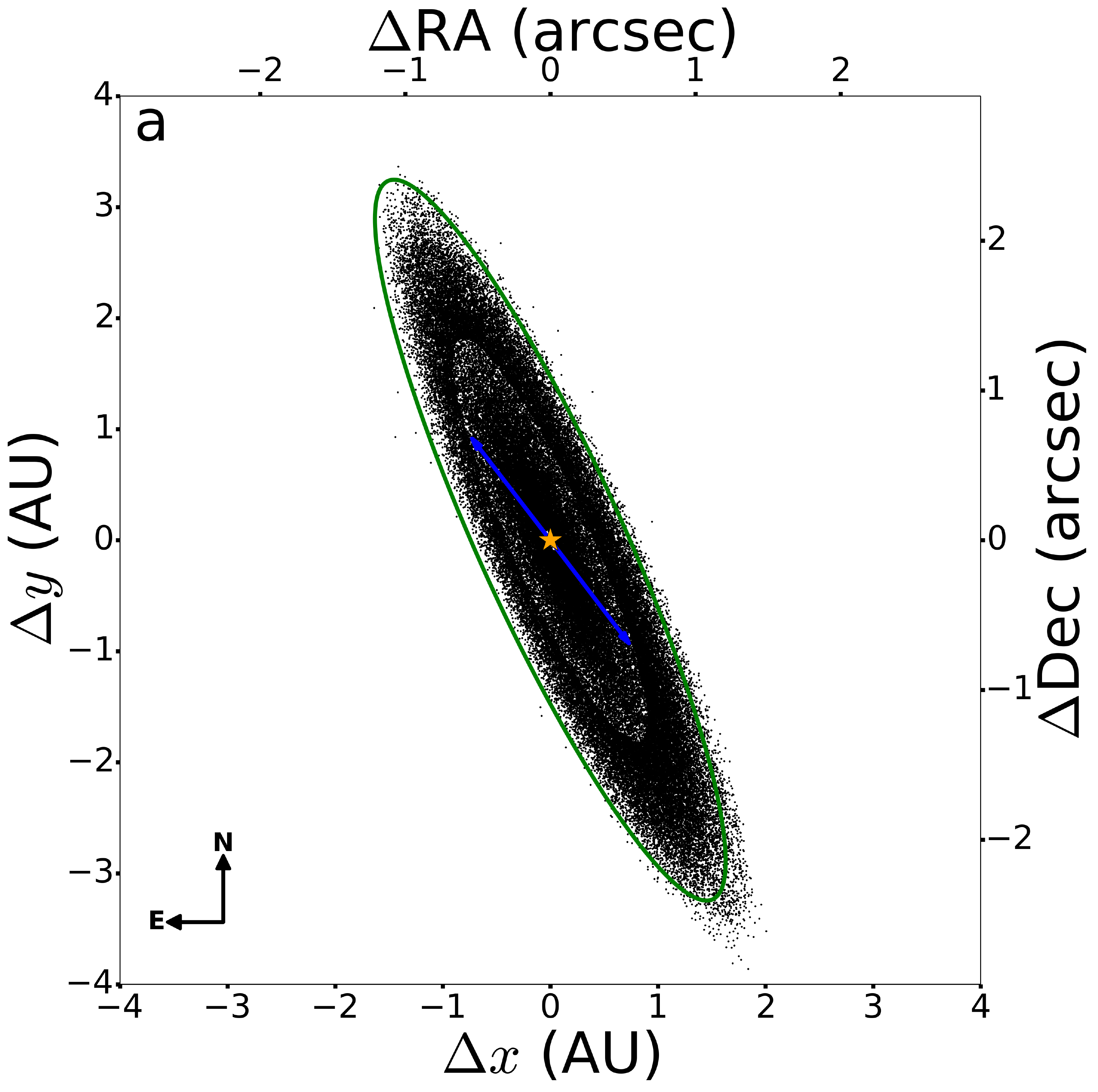}
\includegraphics[scale=0.18]{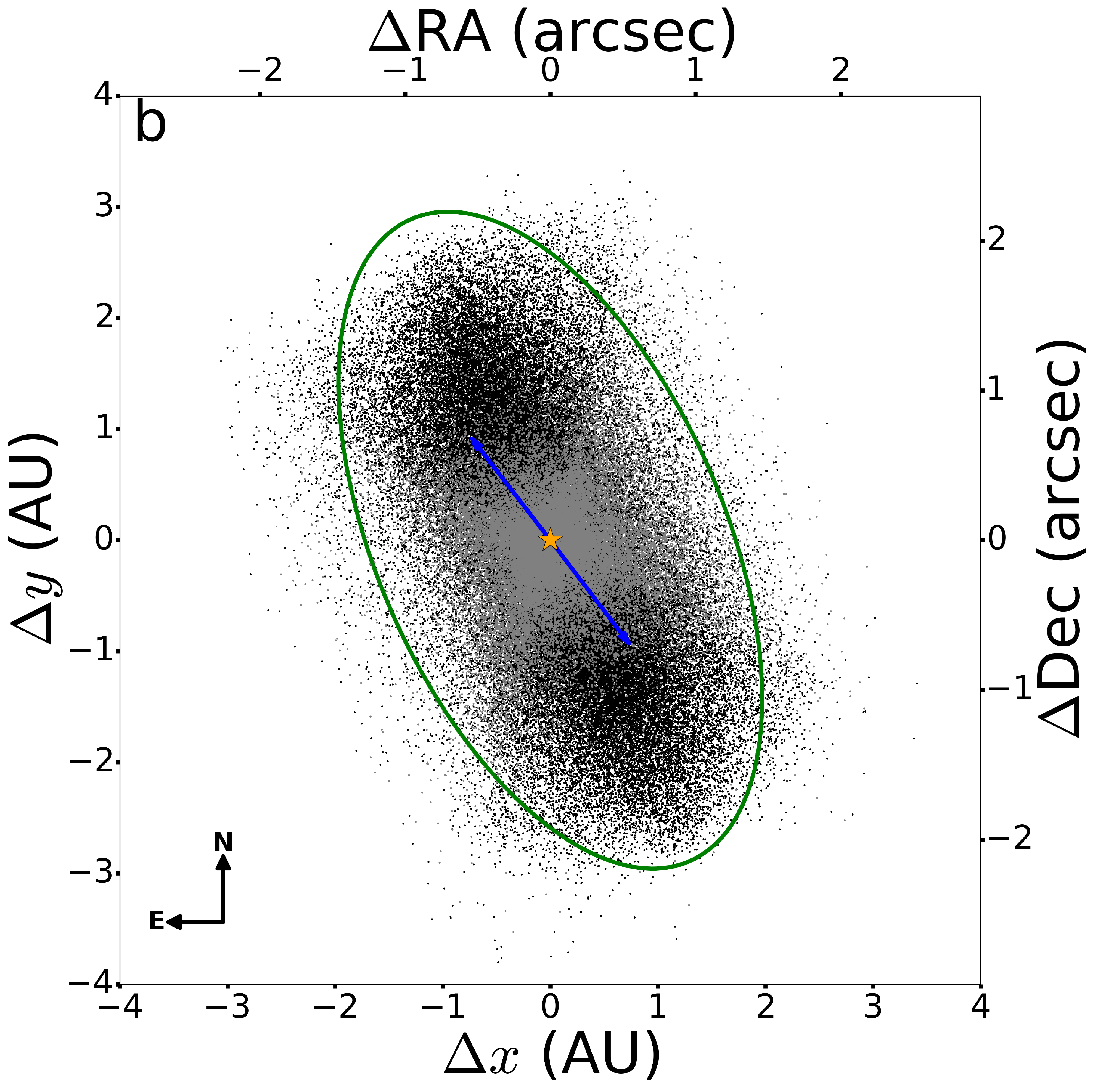}\\
\includegraphics[scale=0.23]{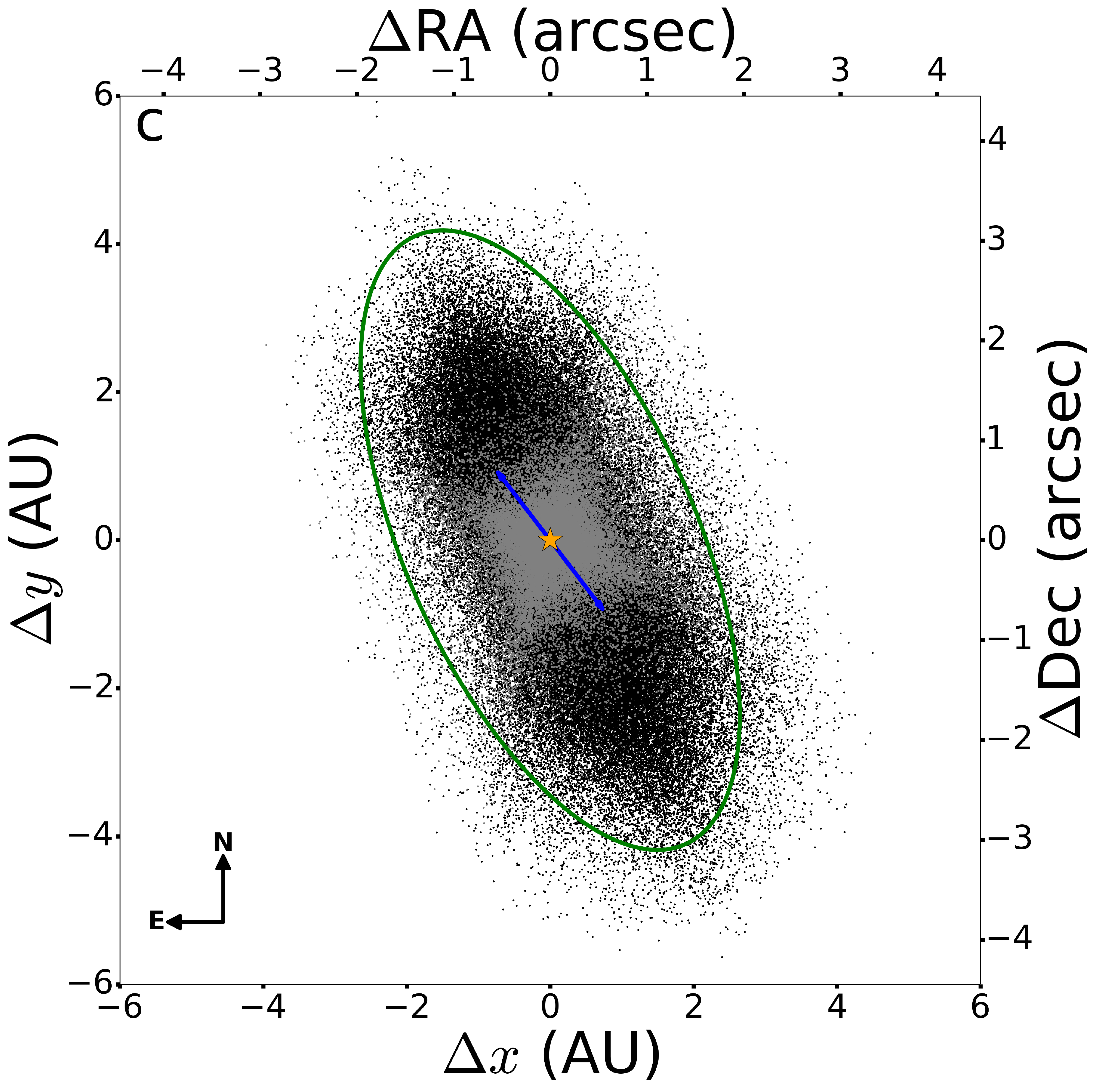}
\caption{Projections on the sky plane of those test particles centered on $\alpha$ Cen A that survived 10 Myr for the initial eccentric (a), prograde inclined (b), and retrograde inclined runs (c).  These results have been summed at approximately the same binary phase (apastron) over 10 binary orbits in order view particles at various phases in their orbits and thereby increase the effective number of test particles simulated.  For each distribution, we also show an ellipse centered on the host star (green) that corresponds to the area spanned by $>$99\% of the test particles and is determined by the covariance in the statistical distribution.  For the eccentric runs (a), the semimajor axis of the green ellipse is $\sim$3.5 AU and represents a larger radial extent along the binary orbit on the sky than the prograde inclined runs (b).  For the inclined runs, the test particles have been separated by color (gray particles plotted on top of black) for those that are in the Lidov-Kozai regime (gray) and those that are not (black).  We also note that the retrograde inclined runs (c) extend farther along the binary orbit and higher in altitude relative to the sky projected binary plane as compared to the prograde inclined runs (b).  The scale for these results are given in distances projected on the sky in AU (left and bottom axes) and arcseconds (top and right axes). \label{sky_part}}
\end{figure*}

We compare the size, location and shape of the distribution of stable particles on inclined prograde orbits about $\alpha$ Cen A at different phases of the binary orbit.  The parameters given in Table \ref{tab:phase} represent different ellipses determined through the statistical covariance of the test particles for the prograde, inclined simulations.  We produce statistical distributions of the test particles by co-adding them across 10 consecutive binary periods at each of 16 different phases.  This process of co-adding helps fill out the area on the sky for which planets might occupy without infringing upon structures introduced by the binary interaction.  We then compute the covariances that encloses $\sim$99\% of the particles on the sky plane for a given phase (or mean anomaly) of the binary.  The results in Table \ref{tab:phase} show that the parameters of the distribution do not vary that much across phases of the binary orbit.  Thus we give the co-added distributions across 10 binary periods but sampled at a single binary phase, apastron, in Figure \ref{sky_part}.  Figure \ref{sky_part} shows views centered on $\alpha$ Cen A using our resulting distribution of stable particles after 10 Myr of their evolution in the eccentric (Fig. \ref{sky_part}a), prograde (Fig. \ref{sky_part}b), and retrograde (Fig. \ref{sky_part}c) runs.      

Figure \ref{sky_part}a shows the distribution of particles along with a green ellipse with a semimajor axis $\lambda_x$ and semiminor axis $\lambda_y$ from Table \ref{tab:phase} at $\sim$180$^\circ$.  We find some structure imposed by the binary orbit onto each distribution through an enhancement (region where stability is more favored) of points on the side between the stars when they are at apastron.  Finally, the eccentric runs begin planar with the binary  (and don't evolve significantly in inclination), so that the resulting distribution of particles looks like a tilted disk when viewed on the sky.  

Figures \ref{sky_part}b and \ref{sky_part}c follow the same procedure but use the results from the prograde and retrograde runs, respectively.  These results are naturally more inclined than those in Fig. \ref{sky_part}a and hence they extend further from the binary plane.  On the sky plane this extension increases in the NorthWest and SouthEast directions.  In Figure \ref{sky_part}b, we further decompose these results and show those (in black) with initial mutual inclination below the critical value for the L-K mechanism as a distinct population relative to the distribution (gray) within the L-K regime.  These two regimes cover very different areas on the sky, but some ambiguity can exist for orbits $\sim$1 AU from the host star because they can occupy the same area on the sky. Figure \ref{sky_part}c shows the results using the retrograde runs.  In this case a much larger portion of the sky can be covered, with the area enclosed by the green ellipse  increasing by $\sim$50\%. 

\section{Conclusions}
\begin{figure*}
{
\plotone{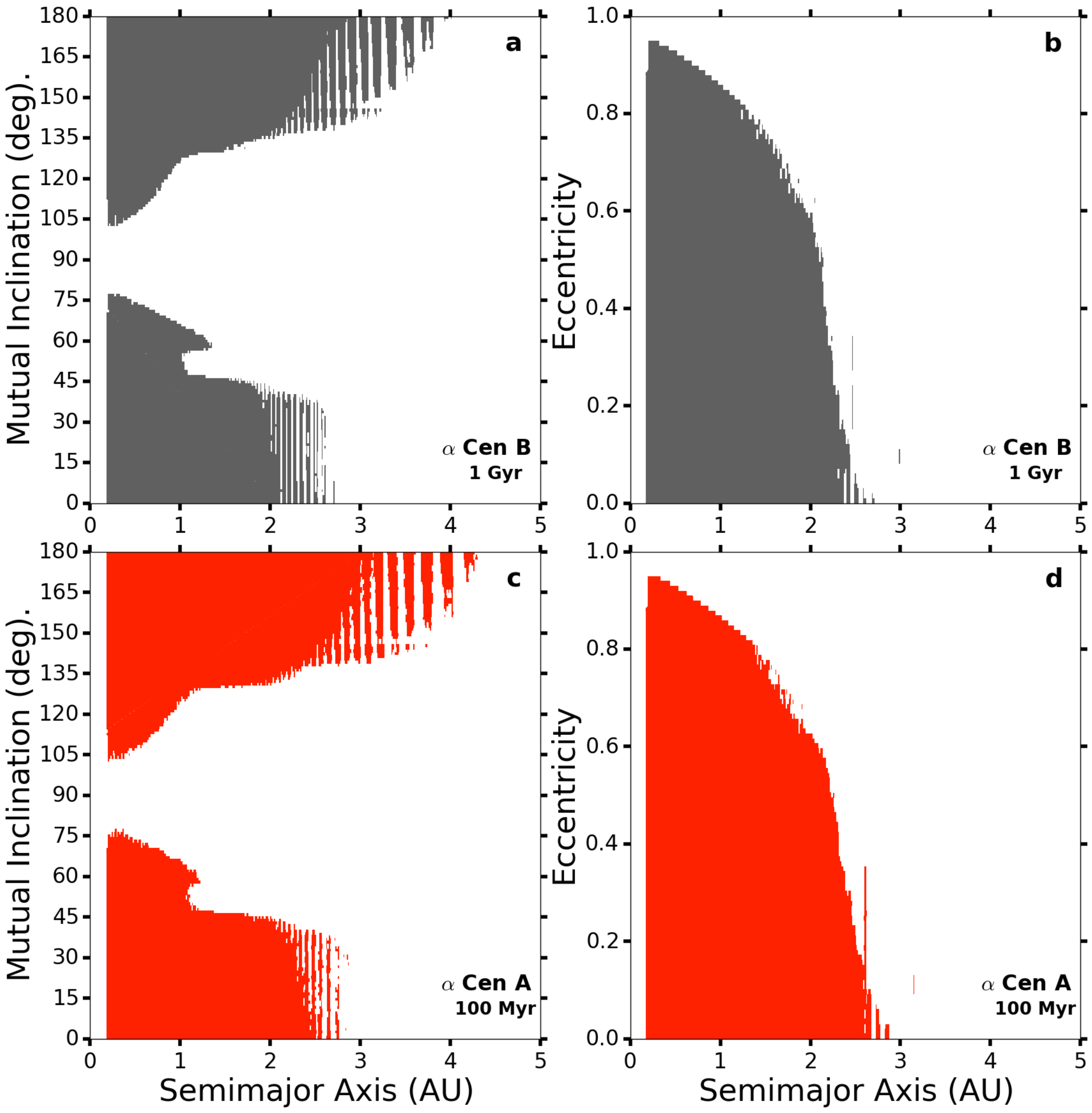}}
\caption{Initial conditions for particles on circumstellar orbits that survived for the entire duration of our simulations.  The four panels represent: (a) circular orbits about $\alpha$ Cen B; (b) orbits about $\alpha$ Cen B that are prograde and lie in the plane of the binary; (c) circular orbits about $\alpha$ Cen A; (d) orbits about $\alpha$ Cen A that are prograde and lie in the plane of the binary.  Particles orbiting near the stability boundary were integrated for 1 Gyr around $\alpha$ Cen B and for 100 Myr around $\alpha$ Cen A; those in the very stable regions close to their star and with low eccentricity and inclination were only integrated for 10 Myr (see Figs. 1 and 4 for details). 
 \label{aCen_sum_alt}}
\end{figure*}

Our simulations show that circumstellar planets (test particles), within the habitable zone of either $\alpha$ Cen A or $\alpha$ Cen B, remain in circumstellar orbit even with moderately high values of initial eccentricity or mutual inclination relative to the binary orbital plane ({Figure \ref{aCen_sum_alt}}).  As a consequence of stability, we find that the dynamical interactions shape the area of the sky planet where we might look for planets.  We confirm the findings of WH97 that particles on circumstellar orbits that are highly inclined relative to the orbital planet of the binary tend to be lost quickly and that retrograde orbits are more stable than prograde ones.  We find that the removal of test particles from near 2.5 AU is not as severe as previously estimated and that the stability boundary for initially eccentric planets can be approximated with a linear function in the periastron of the largest stable semimajor axis.  The realm of stable retrograde orbits is significantly larger than that of prograde orbits on the sky, but planetary formation models suggest that this region is difficult to populate.  

In Figures \ref{aCen_sum_alt}a and \ref{aCen_sum_alt}c , we show that $N$:1 mean motion resonances (MMRs) between the stellar components serve to remove test bodies at their respective locations between 2 -- 3 AU from the host star, likely via resonance overlap \citep{Mudryk2006} in a similar fashion as the Kirkwood gaps of the Asteroid Belt \citep{Wisdom1980}.  This behavior has been shown to be likely in binary star systems with small secondary to primary mass ratios (e.g., \citet[$\gamma$ Cephei]{Satyal2013}, \citet[HD 196885]{Satyal2014}) through the use of the Mean Exponential Growth of Nearby Orbits (MEGNO) chaos indicator \citep{Cincotta2000,Gozdziewski2001}.  Within the regime of critical inclination $\sim$40$^\circ-140^\circ$, the Lidov-Kozai (L-K) mechanism \citep{Lidov1962,Kozai1962} works efficiently to limit the potential stability \citep{Innanen1997} and is somewhat asymmetric about the division (90$^\circ$) between prograde and retrograde.  However, our simulations of eccentric, planar test bodies show both gaps and regions of increased stability due to the $N$:1 MMRs, which likely depends upon the initial phase of the test bodies.  {(Note that the regions of enhanced stability at the 15:1 and 20:1 resonances extend over more than one value of semimajor axis, so the differences seen between these and neighboring resonances cannot be the result of our grid just hitting the ``right'' values of semimajor axis for these mean motion resonances.)}  A Solar System analog to these lucky test particles are the Hilda asteroids, which are near perihelion when they have conjunctions with Jupiter thereby avoiding destabilizing close encounters.

\begin{figure*}
{
\plotone{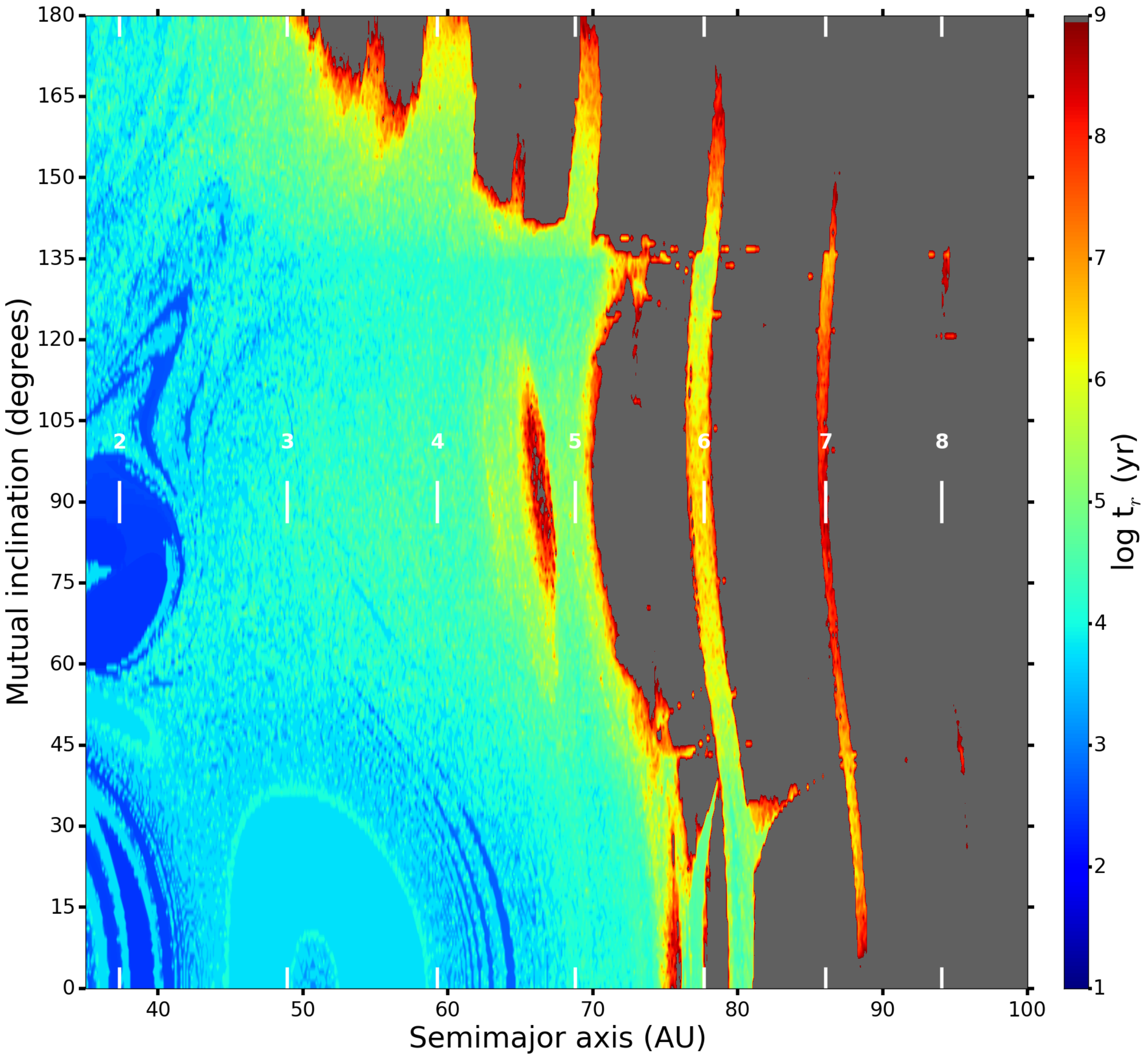}}
\caption{Similar to Figure \ref{aCen_ia_time} but these runs consider circumbinary orbits with initial semimajor axis (relative to the center of mass of the binary) in the 35 -- 100 AU range.  The (external) $N$:1 mean motion resonances are indicated by tick marks, and the color scale representing the removal (ejection or collision) time of the test particles is the same as in Fig. \ref{aCen_ia_time}.  \label{CBP_time}}
\end{figure*}

We also analyze the regions of parameter space that allow for stability of circumbinary planets.  In contrast to the prograde circumstellar case, greater stability is achieved near polar ($i_o\approx 90^\circ$) orbits rather than low inclination ($i_o\approx 0^\circ$) ones (Figure \ref{CBP_time}).  The MMRs also introduce instability at select intervals, but at such great distances from the stellar components that some retrograde configurations ($i_o>165^\circ$) can overcome these effects and survive for long timescales.  Our results and WH97 agree that the retrograde regime provides slightly more stable regions of parameter space than in prograde and that broad regions of stability exist beyond $\sim$80 AU independent of the mutual inclination assumed.

\acknowledgments
B. Q. gratefully acknowledges support by an appointment to the NASA Postdoctoral Program at
the Ames Research Center, administered by Oak Ridge Associated Universities through a contract with NASA. The authors
thank Ruslan Belikov for stimulating conversations over the course of this work. We thank R. Belikov, A. Dobrovolskis,
and E. Quintana for helpful comments on the manuscript.  This work was supported in part by NASA's Astrophysics Research and Analysis program under Proposal No. 13-APRA13-0178.




\bibliographystyle{aasjournal}
\bibliography{refs}

\begin{thebibliography}{}
\expandafter\ifx\csname natexlab\endcsname\relax\def\natexlab#1{#1}\fi

\bibitem[{{Belikov} {et~al.}(2015){Belikov}, {ACEND Team}, \& {ACESat
  Team}}]{Belikov2015a}
{Belikov}, R., {ACEND Team}, \& {ACESat Team}. 2015, in American Astronomical
  Society Meeting Abstracts, Vol. 225, American Astronomical Society Meeting
  Abstracts, 311.01

\bibitem[{Belikov {et~al.}(2015)Belikov, Bendek, Thomas, Males, \&
  Lozi}]{Belikov2015b}
Belikov, R., Bendek, E., Thomas, S., Males, J., \& Lozi, J. 2015, in Techniques
  and Instrumentation for Detection of Exoplanets {VII}, ed. S.~Shaklan
  ({SPIE}-Intl Soc Optical Eng)

\bibitem[{Bendek {et~al.}(2015)Bendek, Belikov, Lozi, Thomas, Males, Weston, \&
  McElwain}]{Bendek2015}
Bendek, E., Belikov, R., Lozi, J., {et~al.} 2015, in Techniques and
  Instrumentation for Detection of Exoplanets {VII}, ed. S.~Shaklan
  ({SPIE}-Intl Soc Optical Eng)

\bibitem[{{Benest}(1988)}]{Benest1988}
{Benest}, D. 1988, \aap, 206, 143

\bibitem[{{Chambers} {et~al.}(2002){Chambers}, {Quintana}, {Duncan}, \&
  {Lissauer}}]{Chambers2002}
{Chambers}, J.~E., {Quintana}, E.~V., {Duncan}, M.~J., \& {Lissauer}, J.~J.
  2002, \aj, 123, 2884

\bibitem[{{Chirikov}(1979)}]{Chirikov1979}
{Chirikov}, B.~V. 1979, \physrep, 52, 263

\bibitem[{{Cincotta} \& {Sim{\'o}}(2000)}]{Cincotta2000}
{Cincotta}, P.~M., \& {Sim{\'o}}, C. 2000, \aaps, 147, 205

\bibitem[{{{\'C}uk} \& {Gladman}(2005)}]{Cuk2005}
{{\'C}uk}, M., \& {Gladman}, B.~J. 2005, \apjl, 626, L113

\bibitem[{{Deitrick} {et~al.}(2015){Deitrick}, {Barnes}, {McArthur}, {Quinn},
  {Luger}, {Antonsen}, \& {Benedict}}]{Deitrick2015}
{Deitrick}, R., {Barnes}, R., {McArthur}, B., {et~al.} 2015, \apj, 798, 46

\bibitem[{{Demory} {et~al.}(2015){Demory}, {Ehrenreich}, {Queloz}, {Seager},
  {Gilliland}, {Chaplin}, {Proffitt}, {Gillon}, {G{\"u}nther}, {Benneke},
  {Dumusque}, {Lovis}, {Pepe}, {S{\'e}gransan}, {Triaud}, \&
  {Udry}}]{Demory2015}
{Demory}, B.-O., {Ehrenreich}, D., {Queloz}, D., {et~al.} 2015, \mnras, 450,
  2043

\bibitem[{{Dumusque} {et~al.}(2012){Dumusque}, {Pepe}, {Lovis},
  {S{\'e}gransan}, {Sahlmann}, {Benz}, {Bouchy}, {Mayor}, {Queloz}, {Santos},
  \& {Udry}}]{Dumusque2012}
{Dumusque}, X., {Pepe}, F., {Lovis}, C., {et~al.} 2012, Nature, 491, 207

\bibitem[{{Dvorak}(1986)}]{Dvorak1986}
{Dvorak}, R. 1986, \aap, 167, 379

\bibitem[{{Endl} {et~al.}(2015){Endl}, {Bergmann}, {Hearnshaw}, {Barnes},
  {Wittenmyer}, {Ramm}, {Kilmartin}, {Gunn}, \& {Brogt}}]{Endl2015}
{Endl}, M., {Bergmann}, C., {Hearnshaw}, J., {et~al.} 2015, International
  Journal of Astrobiology, 14, 305

\bibitem[{{Foreman-Mackey} {et~al.}(2013){Foreman-Mackey}, {Hogg}, {Lang}, \&
  {Goodman}}]{Foreman-Mackey2013}
{Foreman-Mackey}, D., {Hogg}, D.~W., {Lang}, D., \& {Goodman}, J. 2013, \pasp,
  125, 306

\bibitem[{{Gomes} {et~al.}(2005){Gomes}, {Levison}, {Tsiganis}, \&
  {Morbidelli}}]{Gomes2005}
{Gomes}, R., {Levison}, H.~F., {Tsiganis}, K., \& {Morbidelli}, A. 2005, \nat,
  435, 466

\bibitem[{{Go{\'z}dziewski} {et~al.}(2001){Go{\'z}dziewski}, {Bois},
  {Maciejewski}, \& {Kiseleva-Eggleton}}]{Gozdziewski2001}
{Go{\'z}dziewski}, K., {Bois}, E., {Maciejewski}, A.~J., \&
  {Kiseleva-Eggleton}, L. 2001, \aap, 378, 569

\bibitem[{{Henon}(1970)}]{Henon1970}
{Henon}, M. 1970, \aap, 9, 24

\bibitem[{{Holman} \& {Wiegert}(1999)}]{Holman1999}
{Holman}, M.~J., \& {Wiegert}, P.~A. 1999, AJ, 117, 621

\bibitem[{{Innanen} {et~al.}(1997){Innanen}, {Zheng}, {Mikkola}, \&
  {Valtonen}}]{Innanen1997}
{Innanen}, K.~A., {Zheng}, J.~Q., {Mikkola}, S., \& {Valtonen}, M.~J. 1997,
  \aj, 113, 1915

\bibitem[{{Jewitt} \& {Haghighipour}(2007)}]{Jewitt2007}
{Jewitt}, D., \& {Haghighipour}, N. 2007, \araa, 45, 261

\bibitem[{{Kozai}(1962)}]{Kozai1962}
{Kozai}, Y. 1962, \aj, 67, 591

\bibitem[{{Lidov}(1962)}]{Lidov1962}
{Lidov}, M.~L. 1962, \planss, 9, 719

\bibitem[{{Lithwick} \& {Naoz}(2011)}]{Lithwick2011}
{Lithwick}, Y., \& {Naoz}, S. 2011, \apj, 742, 94

\bibitem[{{Mudryk} \& {Wu}(2006)}]{Mudryk2006}
{Mudryk}, L.~R., \& {Wu}, Y. 2006, \apj, 639, 423

\bibitem[{{Naoz} {et~al.}(2011){Naoz}, {Farr}, {Lithwick}, {Rasio}, \&
  {Teyssandier}}]{Naoz2011}
{Naoz}, S., {Farr}, W.~M., {Lithwick}, Y., {Rasio}, F.~A., \& {Teyssandier}, J.
  2011, \nat, 473, 187

\bibitem[{{Plavchan} {et~al.}(2015){Plavchan}, {Chen}, \&
  {Pohl}}]{Plavchan2015}
{Plavchan}, P., {Chen}, X., \& {Pohl}, G. 2015, \apj, 805, 174

\bibitem[{{Popova} \& {Shevchenko}(2012)}]{Popova2012}
{Popova}, E.~A., \& {Shevchenko}, I.~I. 2012, Astronomy Letters, 38, 581

\bibitem[{{Pourbaix} {et~al.}(2002){Pourbaix}, {Nidever}, {McCarthy}, {Butler},
  {Tinney}, {Marcy}, {Jones}, {Penny}, {Carter}, {Bouchy}, {Pepe}, {Hearnshaw},
  {Skuljan}, {Ramm}, \& {Kent}}]{Pourbaix2002}
{Pourbaix}, D., {Nidever}, D., {McCarthy}, C., {et~al.} 2002, \aap, 386, 280

\bibitem[{{Quintana} {et~al.}(2007){Quintana}, {Adams}, {Lissauer}, \&
  {Chambers}}]{Quintana2007}
{Quintana}, E.~V., {Adams}, F.~C., {Lissauer}, J.~J., \& {Chambers}, J.~E.
  2007, ApJ, 660, 807

\bibitem[{{Quintana} {et~al.}(2002){Quintana}, {Lissauer}, {Chambers}, \&
  {Duncan}}]{Quintana2002}
{Quintana}, E.~V., {Lissauer}, J.~J., {Chambers}, J.~E., \& {Duncan}, M.~J.
  2002, ApJ, 576, 982

\bibitem[{{Rabl} \& {Dvorak}(1988)}]{Rabl1988}
{Rabl}, G., \& {Dvorak}, R. 1988, \aap, 191, 385

\bibitem[{{Satyal} {et~al.}(2014){Satyal}, {Hinse}, {Quarles}, \&
  {Noyola}}]{Satyal2014}
{Satyal}, S., {Hinse}, T.~C., {Quarles}, B., \& {Noyola}, J.~P. 2014, \mnras,
  443, 1310

\bibitem[{{Satyal} {et~al.}(2013){Satyal}, {Quarles}, \& {Hinse}}]{Satyal2013}
{Satyal}, S., {Quarles}, B., \& {Hinse}, T.~C. 2013, \mnras, 433, 2215

\bibitem[{{Th{\'e}bault} \& {Haghighipour}(2014)}]{Thebault2014}
{Th{\'e}bault}, P., \& {Haghighipour}, N. 2014, in Planetary Exploration and
  Science: Recent Results and Advances, ed. S.~Jin, N.~Haghighipour, \& W.~Ip
  (Berlin Heidelberg: Springer), 309

\bibitem[{{Th{\'e}bault} {et~al.}(2008){Th{\'e}bault}, {Marzari}, \&
  {Scholl}}]{Thebault2008}
{Th{\'e}bault}, P., {Marzari}, F., \& {Scholl}, H. 2008, MNRAS, 388, 1528

\bibitem[{{Th{\'e}bault} {et~al.}(2009){Th{\'e}bault}, {Marzari}, \&
  {Scholl}}]{Thebault2009}
---. 2009, MNRAS, 393, L21

\bibitem[{{Tsiganis} {et~al.}(2005){Tsiganis}, {Gomes}, {Morbidelli}, \&
  {Levison}}]{Tsiganis2005}
{Tsiganis}, K., {Gomes}, R., {Morbidelli}, A., \& {Levison}, H.~F. 2005, \nat,
  435, 459

\bibitem[{{Wiegert} \& {Holman}(1997)}]{Wiegert1997}
{Wiegert}, P.~A., \& {Holman}, M.~J. 1997, AJ, 113, 1445, (WH97)

\bibitem[{{Wisdom}(1980)}]{Wisdom1980}
{Wisdom}, J. 1980, \aj, 85, 1122

\end{thebibliography}

\clearpage

\newpage

\newpage

\newpage

\newpage

\begin{deluxetable}{cc}
\tablecolumns{2}
\tablewidth{0pc}
\tablecaption{Starting Orbital Elements of the Binary Stars}
\tablehead{\colhead{Element} & \colhead{Value}} 

\startdata
$a$($\prime\prime$) & 17.57 $\pm$ 0.022 \\
$a$(AU)$^*$ & 23.52 $\pm$ 0.036 \\
$i$($^\circ$) & 79.20 $\pm$ 0.041 \\
$\omega$($^\circ$) & 231.65 $\pm$ 0.076 \\
$\Omega$($^\circ$) & 204.85 $\pm$ 0.084 \\
$e$ & 0.5179 $\pm$ 0.00076 \\
P(yr) & 79.91 $\pm$ 0.011 \\
M$_{\rm A}$(M$_\odot$) & 1.105 $\pm$ 0.0070 \\
M$_{\rm B}$(M$_\odot$) & 0.934 $\pm$ 0.0061 \\
\enddata

\tablecomments{Orbital ephemeris assumed for the binary orbit taken from \citealt{Pourbaix2002} where the uncertainties in the parameters illustrate the high accuracy of the determined orbital solution.  $^*$The semimajor axis has been derived from other relevant quantities via Kepler's 3$^{\rm rd}$ law.}

\end{deluxetable}

\begin{deluxetable}{cccccc}
\tablecolumns{6}
\tablewidth{0pc}
\tablecaption{Distribution of the Test Particle Survivors}
\tablehead{\colhead{Mean Anom.} & \colhead{$\bar{x}$} & \colhead{$\lambda_{x}$} & \colhead{$\bar{y}$} & \colhead{$\lambda_{y}$} & \colhead{$\theta$} \\ 
\colhead{(deg.)} & \colhead{(AU)} & \colhead{(AU)} & \colhead{(AU)} & \colhead{(AU)} & \colhead{(deg.)}}
\startdata

5.43   & 0.041 & 1.053 & -0.045 & 0.525 & 24.58 \\
72.54  & 0.027 & 1.048 & -0.021 & 0.525 & 24.77 \\
107.56 & 0.034 & 1.042 & -0.041 & 0.529 & 24.54 \\
128.07 & 0.034 & 1.045 & -0.037 & 0.522 & 24.72 \\
142.40  & 0.033 & 1.047 & -0.032 & 0.525 & 25.08 \\
153.70  & 0.035 & 1.052 & -0.036 & 0.527 & 24.51 \\
163.36 & 0.032 & 1.044 & -0.038 & 0.526 & 24.83 \\
172.14 & 0.032 & 1.043 & -0.048 & 0.526 & 24.78 \\
180.55 & 0.032 & 1.042 & -0.037 & 0.526 & 24.67 \\
188.98 & 0.032 & 1.044 & -0.047 & 0.526 & 24.54 \\
197.84 & 0.032 & 1.046 & -0.045 & 0.528 & 24.67 \\
207.66 & 0.032 & 1.043 & -0.036 & 0.527 & 24.72 \\
219.26 & 0.031 & 1.045 & -0.047 & 0.525 & 24.53 \\
234.17 & 0.031 & 1.043 & -0.037 & 0.526 & 24.56 \\
255.90  & 0.037 & 1.048 & -0.048 & 0.527 & 24.84 \\
294.04 & 0.031 & 1.049 & -0.044 & 0.526 & 24.73 \\
\enddata

\tablecomments{Statistical properties in the distribution of test particle survivors (prograde, inclined around $\alpha$ Cen A) on the sky co-added over 10 binary orbits at 16 different mean anomalies that are equally spaced in time. The values in $\bar{x},\bar{y}$ represent the centers of each distribution and the values $\lambda_x,\lambda_y$ are the eigenvalues of the covariance matrix from the respective distributions. The angle $\theta$ corresponds to the angle of the largest eigenvector relative to the positive $y$-axis of the sky coordinates.} 
\label{tab:phase}
\end{deluxetable}

%
%

\end{document}